\documentclass[]{spie}  

\usepackage{amsmath,amsfonts,amssymb}
\usepackage{graphicx}
\usepackage[colorlinks=true, linkcolor=blue, filecolor=blue, citecolor=blue, urlcolor=blue]{hyperref}    
\usepackage{siunitx}

\title{FAST: A software suite for automatic weather and optical turbulence forecast on ground-based telescope sites}

\author[a]{Turchi, A.}
\author[a]{Masciadri, E.}
\author[a]{Fini, L.}

\affil[a]{INAF-Osservatorio Astrofisico di Arcetri, L.go Enrico Fermi 5, Firenze, Italy}

\authorinfo{Send correspondence to Alessio Turchi - e-mail: alessio.turchi@inaf.it}

\begin{document} 
\maketitle

\begin{abstract}
In this contribution we present the FAST, which is a comprehensive software suite that aims to streamline and automatically manage the forecast of atmospheric and astroclimatic parameters (provided respectively by Meso-Nh and Astro-Meso-Nh models) on large ground-based telescope installations. The forecast of the aforementioned parameters is becoming crucial for the operation of the large telescope installations which possess atmospheric-sensitive equipment equipped with Adaptive Optics (AO) systems. FAST performs automatically all the steps of an atmosphere forecast process: initialisation and forcing data, atmospheric simulation, postprocessing and managing of the outputs.The role of such service is useful both in optimizing beforehand AO instruments to the next atmospheric conditions and in planning telescope observations (especially in “service mode”) in order to maximize the scientific output. FAST was applied first to the ALTA Center project (3), which provides forecasts for the LBT telescope. Then it was extended to the more recent project FATE that is a similar forecast system applied to the VLT. Since its first version FAST evolved and it has has been modified to fit with the different technical specifications of the different projects gaining in modularity. It is now able to provide forecasts on different timescales (from days to hours before) and to provide forecast during night and day time. After several years of continuous development we can say that FAST reached full maturity and it is now ready for applications to other projects/sites.
\end{abstract}


\section{INTRODUCTION}
\label{sec:intro} 
Accurate forecasting of specific atmospheric parameters, including Optical Turbulence (OT), has become increasingly essential for ground-based astronomy, especially as instrumentation advances towards the era of Extremely Large Telescopes (ELTs). OT is a critical factor limiting the attainment of high angular resolution in ground-based observations, and atmospheric conditions in general significantly affect observations with top-tier telescopes. For instance, seeing ($\epsilon_0$) conditions affects the attainable angular resolution from all ground-based observations, however it must be noted that many atmospheric parameters other than OT play a important role on observations quality, e.g. Precipitable Water Vapor (PWV) can constrain infrared observations  [\citenum{pwv}]. 

To mitigate the effects of OT, Adaptive Optics (AO) systems have been developed over the past few decades. AO an correct many of the wavefront distortions induced by OT, particularly in high-contrast imaging systems. For example, AO systems on 8-10m class telescopes can achieve a Strehl Ratio (SR) of up to 90\% in the H band under favorable conditions (seeing of around 0.4”). However, the performance of these systems is still dependent on the prevailing atmospheric and OT conditions, achieving peak correction only under ideal conditions, which are infrequent. These conditions are characterized by the $\epsilon_0$ parameter, isoplanatic angle ($\theta_0$), and wavefront coherence time ($\tau_0$) [\citenum{OT1}, \citenum{OT2}, \citenum{OT3}].

The increasing complexity of Wide Field Adaptive Optics (WFAO) systems means that atmospheric conditions will have an even greater impact on the scientific output of advanced telescopes in the future. Consequently, forecasting atmospheric and OT conditions is crucial for optimizing telescope schedules and ensuring that observations are conducted under conditions that maximize scientific returns [\citenum{scheduling}, \citenum{SM}]. The ability to select scientific programs and configure instrumentation based on atmospheric conditions, is known as 'flexible scheduling' or 'Service Mode'. This scheduling, which takes into account both atmospheric conditions and the quality of scientific programs, is becoming a standard practice for these observatories.

OT forecasting is an extremely challenging goal with significant impacts across various fields. As we highlighted before it is crucial in ground-based astronomy, and beyond astronomy, OT forecasting is vital for applications involving wavefront propagation through atmospheric turbulence at short wavelengths, such as optical communications between space and Earth in the visible or near-infrared spectrum [\citenum{space}]. These applications promise significantly faster data transmission and better security compared to radio signals.

The use of mesoscale non-hydrostatic models to reconstruct OT profiles was initially proposed by Masciadri et al. (1999) [\citenum{astroMNH}], employing the prognostic equation of turbulent kinetic energy (TKE). This approach has been validated and refined through numerous studies using the Astro-Meso-NH model developed by the same authors, proving effective in forecasting OT and integrated astroclimatic parameters for astronomical applications. Recent developments have also seen other mesoscale models and General Circulation Models (GCMs) being used [\citenum{GCM1}, \citenum{GCM2}], though mesoscale models generally offer better performance in estimating seeing and other atmospheric parameters [\citenum{GCMcon}].\\

In recent years, in order to improve the forecast performance at short time scales (up to few hours in the future), we developed new methods to make use of the on-site measurements of the atmospheric and OT parameters in order to enhance the astro-MNH model predictions [\citenum{shortTS}]. This drastically improved the prediction accuracy for short-term telescope planning, however it increased the overall complexity of the forecast system.\\

In this article, we present FAST, which is an acronym for ``Forecast Automation System for Telescopes'', a versatile software suite developed to streamline and automate the prediction of atmospheric and OT parameters for large ground-based telescope installations. These predictions are generated by the Meso-Nh [\citenum{MNH1}, \citenum{MNH2}] and Astro-Meso-Nh [\citenum{astroMNH}] models, but also include a module to run also forecast routines such as AutoRegressive methods (AR) or Machine Learning (ML). FAST handles the entire atmospheric forecasting workflow, including data initialization, atmospheric simulation, post-processing, and output management. This automated service is crucial for integration with the telescope planning instruemnts and allow for an efficient planning of the telescope observations.

FAST was first implemented in the ALTA Center project [\citenum{predLBT1}, \citenum{predLBT2}], providing forecasts for the LBT telescope, and was later extended to the FATE project for the VLT [\citenum{predVLT},\citenum{pwv}]. Over time, FAST has evolved to meet the specific technical requirements of various projects, enhancing its modularity. It now delivers forecasts over different timescales, ranging from days to hours in advance, and supports both nighttime and daytime operations. After years of ongoing development, FAST has reached a mature state and is now prepared for deployment in other projects and locations.

\section{The FAST workflow}
\label{sec:fast}

\subsection{Meso-NH}
\label{sub:mnh}
While this contribution is focused only on the FAST software, we give here a short summary of the Meso-NH simulation process which is managed by FAST itself. For more details we suggest to read the other papers from the same authors which cover the simulation process more in detail on the specific case studies [\citenum{SM}, \citenum{astroMNH}, \citenum{operational}, \citenum{predLBT2}].\\

The atmospheric simulation that provide the forecast of the relevant parameters, such as $\epsilon_0$, $\tau_0$, $theta_0$, PWV, wind speed and direction, temperature, relative humidity, etc..., is based on the Meso-NH software, which is developed by the Laboratorie d'Aérologie in Tolouse (France), together with Meteo France [\citenum{MNH1}, \citenum{MNH2}]. While Meso-NH produces detailed mesoscale simulations of most of the atmospheric parameters, the Astro-Meso-Nh code, which is developed by INAF-OAA [\citenum{astroMNH}], is the Meso-NH module that provides computation of all the OT parameters, included a detailed stratification of the $C_N^2$ profiles.\\
Meso-NH simulations are performed in different steps. The first one is the PGD phase, in which a detailed orographic model of the surface is generated in a region of few hundred kilometers around the relevant point of interest, which in our case is the telescope installation, including al the relevant details about the soil composition and ground cover which are fundamental for a correct description of the ground-atmohspere interaction. The model may use grid-nesting technniques to increase the resolution while approaching the center of the simulated area. The second step is the PREP-REAL phase, which takes the initial atmospheric conditions from a global circulation model (in our case the European Center for Medium Range Weather Forecast - ECMWF) and merge them with the ground model computed in the PGD step thus creating also an initial computational grid. Initial conditions include all the prognostic parameters, such as temperature, pressure, wind speed, and other relevant atmospheric quantities, that are computed starting from satellite data, weather stations and other data gathering instruments scattered all over the globe and then forecasted on a global scale for few days in the future. We take initialization conditions at synoptic hours and use them to initialize the model and also to provide "forcing" during the evolution, i.e. providing mid-point conditions for the atmospheric equations to avoid divergence over a small domain.
The next step is the EXE phase in which the model performs the simulation of the atmosphere for the selected time period and forecast most of the parameters. In an optional subsequent phase, the DIAG phase, few parameters that are not computed in the previous step (such as the $\epsilon_0$ 2D maps and other derived parameters) are computed in a first post-processing phase.

\subsection{The FAST software}

\begin{figure}
\centering
\includegraphics[width=0.9\textwidth]{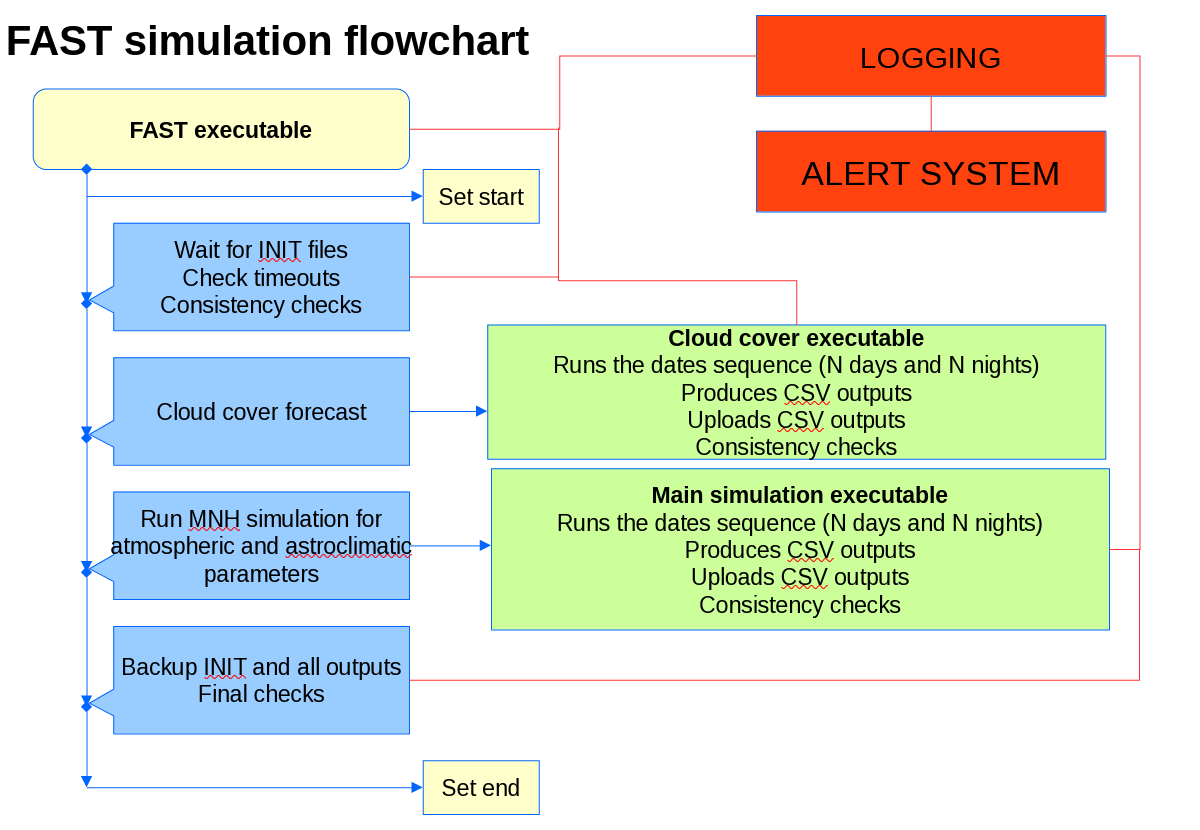}
\caption{Main FAST execution flowchart. This scheme represent the execution of the global FAST workflow that oversees the individual steps of initialization, MNH simulation (standard forecast), backups, logging and alert system.}
\label{fig:fast1}
\end{figure}

\begin{figure}
\centering
\includegraphics[width=0.9\textwidth]{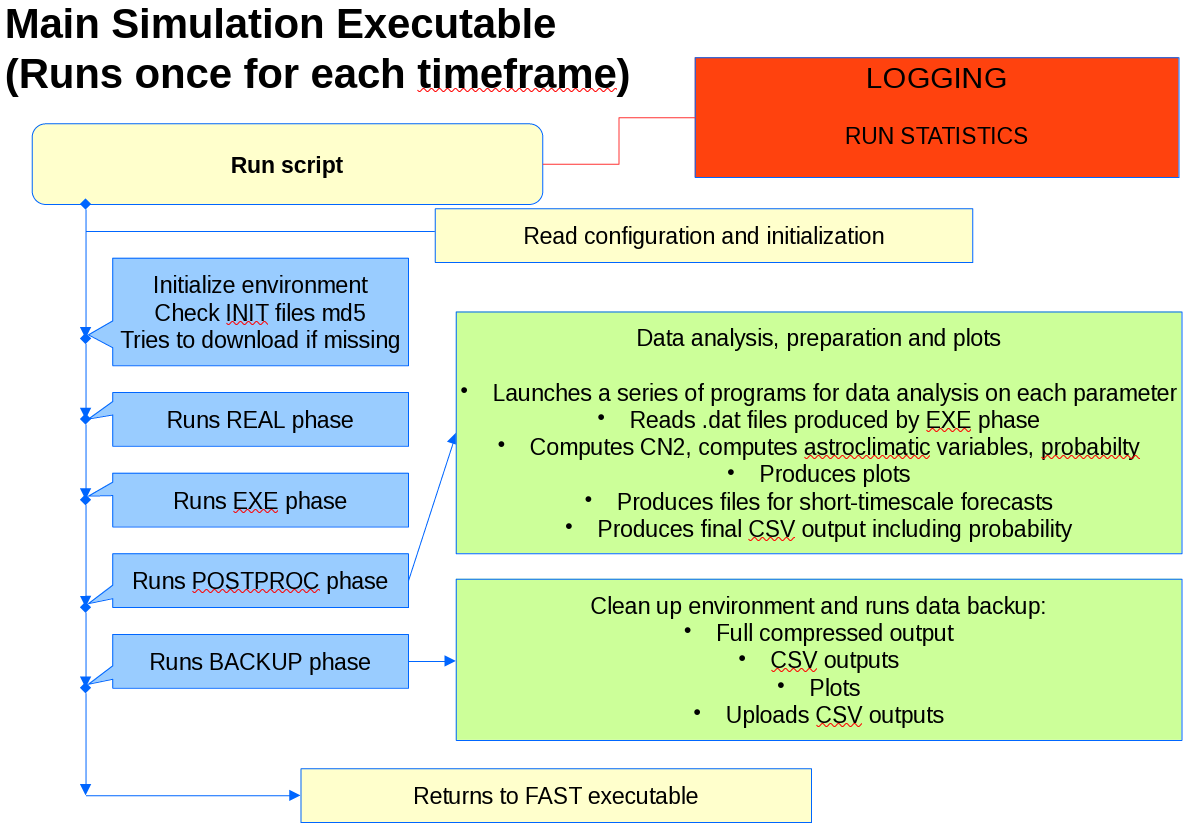}
\caption{This is a detailed scheme of the execution block for each MNH simulation that is run as part of the global workflow. This includes all the simulation steps and the data processing on the output files.}
\label{fig:fast2}
\end{figure}

The FAST software is made up of multiple modules, written mainly in bash and python, while part of the data analysis pipeline is written in fortran. A main process (see Fig, \ref{fig:fast1}) oversee all the individual simulations that may include different time periods, (e.g. differnt nights and/or different days in the future). Typically the simulation process is split between nighttime and daytime because of the different physics involved (which is mostly due to different convective regimes affecting deeply the OT parameters). There is no theoretical limits on how many nights and days can be included in the simulation process, provided that the necessary initialization data is present.\\
A sub-module is responsible for each simulation (see Fig. \ref{fig:fast2}) for each selected timeframe (e.g. night 1, day 1, and so on). This step oversee the individual Meso-NH steps described in sec. \ref{sub:mnh}, recovers the initialization condition from the remote server where ECMWF delivers them and run REAL, EXE and postprocessing procedures. The PGD phase is usually run only once for each telescope site when installing the system, and typically there is no need to re-run it unless the ground changes for some reason.\\
The postprocessing include the optional DIAG phase previously cited and the data analysis of the simulation outputs at high temporal frequency (from few seconds up to two minutes). Here the model produces the final parameters with the desired time frequency and or data processing as requested by the telescope teams and which are then directly delivered to the remote delivery server. This step may also produce graphic plots of the selected variables in order to display them on a website.
A final step is in charge of the backup of the data to a local archive and cleans up the environment for the next simulation.\\

Either the main module and each sub-module has a complex and full set of tests and runtime checks that monitor each step of the automated procedure. This allows to provide deep insight of each failure in order to debug the procedures and also provide accountability for the production of reports for the telescope institution.\\
The whole procedure is tied to a custom logging system that stores all the relevant information to a series of ASCII files on a local and/or remote directory. The logging system is in charge of monitoring the whole procedure (the main module) and the single sub-modules that performs the individual simulations. The logs can also be used to provide statistics on the model behaviour in order to potentially identify unforeseen problems. An alert system is included in the procedure in order to provide fast and detailed information to all the subject involved in the maintenance of the system (this is done via e-mail), who can react in order to solve all the potential problems.\\

The FAST system is organized as a single main executable that runs the sub-modules. Each sub-module is self-contained in a directory and has its own configuration files. A main "globals.var" stores all the global information such as local paths, Meso-NH paths, logging configuration, mail addresses for reporting and all the global simulation parameters. A "pdg.var" file contains all the parameters used to run the PGD phase on the selected spatial domain. A "run.var" file contains all the parameters which are relevant for each Meso-NH simulation, such as number of processors, timestep, nesting configuration, vertical grid and other parameters which are included in the Meso-NH configuration. A "times.var" file include all the information on the timeframes for the selected simulation, such as date, start and end of the forecast, time limits for the analysis, sunset and sunrise times and so on. A "filedef.var" file stores all the information on the initialization files that are used to start the PREP-REAL phase. A final "backup.var" file stores the configuration for data delivery and backup of the simulation.\\
From the above config file, a namelist (the Meso-NH run procedure) is automatically compiled ,starting from a template, for each Meso-NH step and then fed to the model for execution. The model is executed in parallel over the selected hardware architecture.\\

\subsection{Short timescale forecast}

\begin{figure}
\centering
\includegraphics[width=0.9\textwidth]{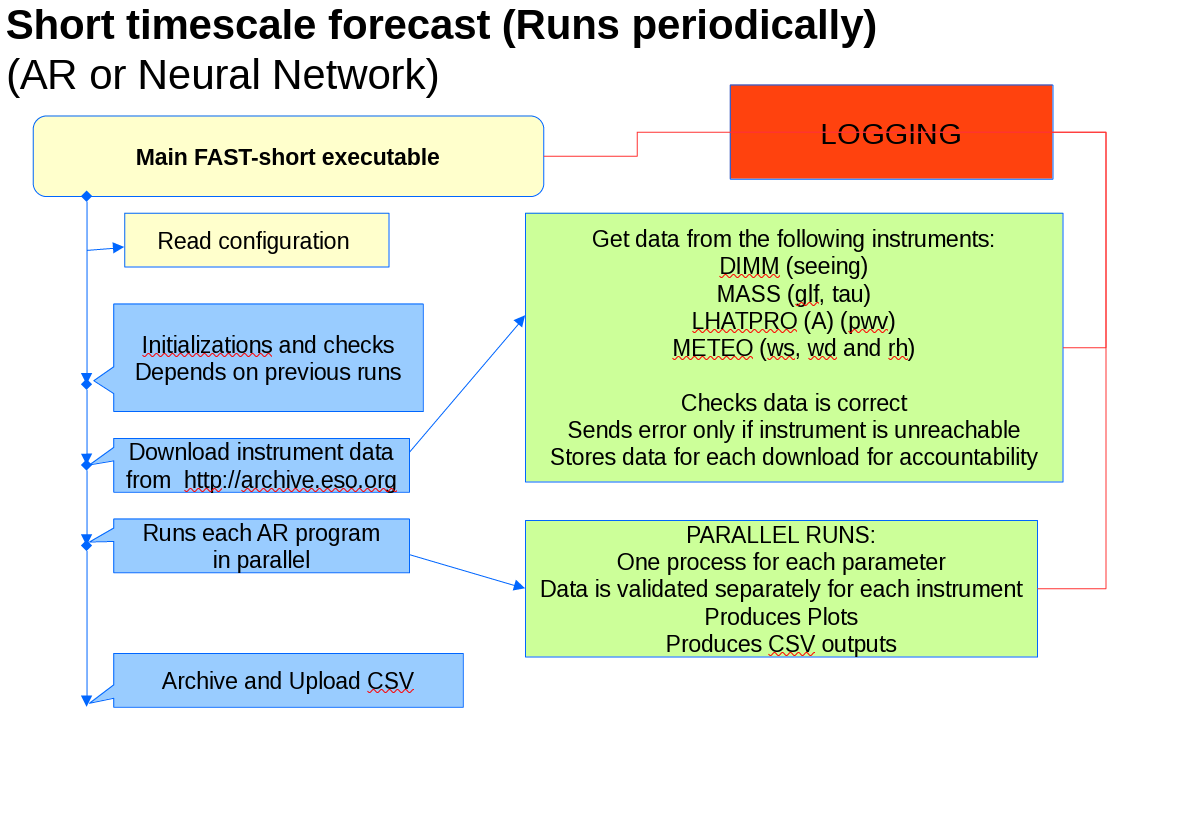}
\caption{This is a scheme of the FAST-short execution, which is run independently of the main FAST workflow and takes chanrge of the short-term forecasts that make use of the measurements from the telescope telemetry streams. This produces the enhanced forecasts for the next few hours in the future.}
\label{fig:fast3}
\end{figure}

During the years, in order to account for the different telescope installations needs, a FAST-short system was developed. This system is in charge of providing a short-term forecast on a reduced timescale (typically few hours in the future) which has however a higher accuracy (see Fig. \ref{fig:fast3}). In this contribution we will not describe the model performances, for which we refer the reader to other articles from the same authors [\citenum{predLBT1}, \citenum{predVLT}].\\
This system is independent of the main FAST executable because it runs on different timescales. Typically once each hour but we are developing it to provide an update of the forecasts every 10 minutes. The logging system is similar but also independent of the main system.\\

FAST-short includes a main executable which runs with the chosen schedule and has a single configuration file. The whole procedure is self-contained in a single directory and is based on a different sub-procedure for each parameter that is included in the short-term forecast, i.e. seeing, PWV or temperature. The system automatically recovers the data from the telescope instruments that measure the specific parameters (i.e. the DIMM for the seeing, the LHATPRO for the PWV or the specific weather monitoring systems for temperature and other atmospheric parameters) by connecting to the telescope telemetry. The system is highly modular and currently it is configure to either connect to the LBT or VLT telescope telemetry data. The real-time instrument data is used to enhance the Meso-NH prediction produced by the previous FAST procedure over the next few hours in the future by making use of a machine-learning procedure.\\

Currently the software uses an autoregressive method (AR) [\citenum{predLBT1}], but it is in the process of beeing enhanced to make use of more complex Neural Network (NN) methods that may further enhance its performance. For this purpose, the system is highly modular and the prediction for each parameter can use a different method. In principle any software can be used to perform each parameter estimation if it can provide the same interface to the main FAST-short executable. Finally all the single forecast processes for each parameter are run in parallel for efficiency.\\

Similarly to the main FAST software, FAST-short is responsible to the whole forecast procedure, from the data retrieval from the telemetry, parallel simulation execution, data delivery to the telescope and final backup of the outputs.\\

\section{Conclusions}
\label{sec:end}
Currently the FAST and FAST-short systems are used to provide daily forecasts for the LBT and VLT telescopes, as part of the ALTA and FATE projects respectively. While FATE started recently this year, ALTA has been active since 2016 with a 99\% or greater uptime and no major failures. FAST software is a valid tool to streamline and automatize the weather forecast process using Meso-NH and Astro-Meso-NH models and is specifically tailored for the needs of large telescope installations. It can provide either long-term forecasts using a mesoscale model or short-term forecasts by enhancing the model forecast with the telescope data telemetry by means of autoregressive or neural network techniques. The software is highly modular and can be easily adapted to different needs. After almost a decade of development the software has reached full maturity and is ready for deployment on other ground-based telescope installations.

\acknowledgments
The authors thanks the Meso-Nh users supporter team who constantly works to maintain the model by developing new packages. This study has been co-funded by the FRCF foundation through the 'Ricerca Scientifica e Tecnologica' action - N.45103, by the contract ENV001 (LBTO), by the contract FATE N. PO102958/ESO/20/95952/FLAB (ESO), by the EU Horizon 2020 research and innovation programme under the grant agreement N. 824135 (SOLARNET).

\end{document}